\PassOptionsToPackage{table}{xcolor}
\documentclass[mlmain]{jmlr}

\usepackage{longtable}

\usepackage{booktabs}
\usepackage[load-configurations=version-1]{siunitx} 
\usepackage{adjustbox}
\usepackage{multirow}
\usepackage{graphicx}

\def\be{\begin{equation}}
\def\ee{\end{equation}}

\theorembodyfont{\upshape}
\theoremheaderfont{\scshape}
\theorempostheader{:}
\theoremsep{\newline}

\jmlrvolume{}
\firstpageno{1}

\jmlryear{2026}
\jmlrworkshop{Symmetry and Geometry in Neural Representations}

\title[Cup and Cap Topological Neural Network]{Cup and Cap Topological Neural Network}

 \author{
 \Name{Marta Niedostatek\nametag{\thanks{Equal contribution.}}} \Email{m.niedostatek@qmul.ac.uk}\\
 \addr Queen Mary University of London
 \AND
 \Name{Ferran {Hernandez Caralt}}\footnotemark[1]\Email{fh455@cam.ac.uk}\\
 \addr University of Cambridge
 \AND
 \Name{Runyue Wang}\Email{runyue.wang@qmul.ac.uk}\\
 \addr Queen Mary University of London
 \AND
 \Name{Federica Baccini}\Email{baccini@diag.uniroma1.it}\\
 \addr Sapienza University of Rome
 \AND
 \Name{Lorenzo Giambagli}\Email{lorenzo.giambagli@fu-berlin.de}\\
 \addr Freie Universität Berlin
 \AND
 \Name{Pietro Li\'o}\Email{pl219@cam.ac.uk}\\
 \addr University of Cambridge
 \AND
 \Name{Ginestra
 Bianconi}\Email{g.bianconi@qmul.ac.uk}\\
 \addr Queen Mary University of London
 }

\begin{document}

\maketitle

\begin{abstract}
Topological Deep Learning (TDL) is designed to learn features associated   not only with nodes but also with edges, triangles and higher
dimensional simplices. However, most TDL approaches, being based on boundary operators and Hodge Laplacians, have the limitation that features and signals cannot be lifted or lowered across more than one dimension per layer. To overcome this limitation, in this work, we propose the adoption of  the cup and cap products. Specifically, we formulate the  Cup and Cap Topological Neural Network (CCNN), a topological deep learning architecture designed to learn node-based variables (or 0-cochains) by taking into account their many-body interactions (e.g. triangles) present in the data. We validate CCNN by assessing its performance on the TopoBench datasets, demonstrating its competitive performance with respect to other simplicial complex neural networks.
\end{abstract}


\section{Introduction}
\label{sec:intro}

To make predictions on relational data, Graph Neural Networks \citep{Scarselli2009,Kipf:2017tc} have long been  very successful tools. They are, however, limited to considering only pairwise relationships in the data. Many real-world systems are higher-order \citep{bianconi2021higher,millan2025topology,Battiston2025} and thus not limited to  pairwise interactions \citep{Battiston2021}. These include social networks \citep{iacopini2019simplicial}, ecological systems \citep{mayfield2017} and biological systems \citep{Gaudelet2018}. To model such complex data, the field of Topological Deep Learning (TDL) \citep{papamarkou2024position,telyatnikov2025,millan2025topology} has been developed and several topological  neural networks \citep{bodnar2021,goh2022,bodnar2022neural,caralt2024joint} have been proposed in the literature.  Simplicial neural networks \citep{ebli2020simplicialneuralnetworks,yang2022,10735150}, which have been inspired by topological dynamics \citep{millan2020explosive,millan2025topology} and topological signal processing \citep{barbarossa2020topological,calmon2023dirac, wang2025dirac,schaub2021}, have emerged as a key component of TDL. Despite the great achievements of the field, which have opened new perspectives for learning features associated with higher-dimensional simplices, many of these architectures, being based on the boundary operators and on the Hodge Laplacians, are not able to lift or lower the learned features across more than one dimension per layer. Thus, for instance, these architectures cannot be used to lift node features up to triangles within one layer.

{\bf Main contributions of the paper}\\
In order to overcome this limitation, in this work we propose the Cup and Cap Topological Neural Network (CCNN). Being based on the cup and cap products, instead of using exclusively the boundary operators, the CCNN allows  lifting and lowering features across an arbitrary number of dimensions per layer. Specifically, the CCNN learns node features by taking into account the higher-order interactions among the nodes. Focusing specifically on the case in which node features are lifted up to triangles, we  test and validate the CCNN across the TopoBench datasets \citep{telyatnikov2025} and we demonstrate the competitiveness of this architecture with respect to existing simplicial complex neural networks.



\section{Background}
\subsection{Simplicial complexes; chains and cochains}
We consider a $d$-dimensional simplicial complex $\mathcal{K}$ with $N_q$ simplices ${\mathbf{e}}_{i_0,i_1,\ldots i_q}$ of dimension $q$ each given by  a set of $q+1$ nodes 
\be
{\mathbf{e}}_{[i_0,i_1,\ldots i_q]}=[i_0,i_1,\ldots, i_q].
\ee
Note that we consider differently oriented simplices as independent basis elements, thus ${\mathbf{e}}_{ij}$ is  linearly independent of ${\mathbf{e}}_{ji}$ and similarly $\hat{\mathbf{e}}_{ij}$ is linearly independent of $\hat{\mathbf{e}}_{ji}$. Although this convention is not typically used in TDL, this convention is allowed and widely used in topology \citep{hatcher2005algebraic}.

We indicate with $Q_q$ the set of $(q+1)$-tuples  $(i_0,i_1,\ldots, i_q)$ corresponding to simplices in $\mathcal{K}$.
The structure of the simplicial complex $\mathcal{K}$ can be described by the set of all its simplices, or alternatively by the set of adjacency tensors ${\mathbf{a}}^{(q)}$ of elements $a^{(q)}_{i_0i_1,\ldots i_q}=1$ if the simplex ${\mathbf{e}}_{i_0,i_1,\ldots, i_q}\in \mathcal{K}$ while otherwise $a^{(q)}_{i_0i_1,\ldots, i_q}=0$.
In algebraic topology, the simplices ${\mathbf{e}}_{i_0,i_1,\ldots ,i_q}$ form the canonical basis for the chains. Specifically, a $q$-chain $\boldsymbol{\theta}\in C_q$ can be expressed as a linear combination of the simplices, i.e.
\be
\boldsymbol{\theta}=\sum_{(i_0,i_1,i_2\ldots{i_q})\in Q_q}\theta_{[i_0,i_1,i_2\ldots i_q]}{\mathbf{e}}_{[i_0,i_1,i_2\ldots i_q]}.
\ee
Interesting examples of $q$-chains are $\boldsymbol{\alpha}^{(q)}\in C_q$, constructed starting from the $q$-adjacency tensors, i.e.
\be
\boldsymbol{\alpha}^{(q)}=\sum_{i_0,i_1,\ldots i_q}a^{(q)}_{i_0 i_1\ldots i_q}{\mathbf{e}}_{[i_0,i_1,\ldots i_q]}.
\ee
A $q$-cochain $\hat{\theta}\in C^q$   is a linear function $\hat{\theta}:C_q\to \mathbb{R}$ associating to each $q$-chain $\boldsymbol{\theta}\in C_q$ a value in $\mathbb{R}$. Indicating with   $\hat{\mathbf{e}}_{[i_0,i_1,\ldots,i_q]}$  the canonical basis for cochains defined on a single simplex ${\mathbf{e}}_{[i_0,i_1,\ldots i_q]}$, with each element $\hat{\mathbf{e}}_{[i_0,i_1,\ldots,i_q]}$ expressing the value that the $q$-cochain takes on the corresponding  simplex,  we can express a $q$-cochain  as  $\hat{\boldsymbol{\theta}}$ given by
\be
\hat{\boldsymbol{\theta}}=\sum_{(i_0,i_1,i_2,\ldots,i_q)\in Q_q}\theta_{[i_0,i_1,i_2,\ldots,i_q]}\hat{{\mathbf{e}}}_{[i_0,i_1,\ldots,i_q]}
\ee
Since  the cochains indicate the signals defined on simplices including nodes, edges, triangles, etc., it follows that they are the natural   mathematical  representation of  the features associated with these simplices.  Therefore, the  general task of Topological Neural Networks (TNNs) is to learn missing elements of cochains defined on simplicial complexes.

\subsection{The boundary and coboundary  operators}
The boundary operator plays a fundamental role in algebraic topology. Although allowing for independent orientation of the simplices changes some of the properties of these operators (see for details \cite{wang2026topology}) the main topological features remain unchanged.
 The boundary operator $\partial_q:C_q\to C_{q-1}$ is a linear operator that maps $q$-chains to $(q-1)$-chains. Since the boundary operator is linear, its action is completely determined by its action on each $q$-simplex ${\mathbf{e}}_{i_0,i_1,\ldots, i_q}$ given by 
\be
\partial_q{{\mathbf{e}}_{[i_0,i_1,\ldots, i_q]}}=\sum_{p=0}^{q}(-1)^p {\mathbf{e}}_{[i_0,i_1,\ldots,i_{p-1},i_{p+1}\ldots i_q]}.
\ee  
From this definition it is clear that the boundary operator maps a $q$-chain formed by a single simplex to a $(q-1)$-chain formed by all the simplices at its boundary, with a well-defined convention of their orientations.
A fundamental topological property is that the {\em boundary of the boundary is null,} i.e. $\partial_{q}\partial_{q+1}=0 \ \forall q\geq 1,$
which implies $\mbox{img }({\partial}_{q+1})\subseteq \mbox{ker }({\partial}_{q})$.
We call a $q$-cycle  a $q$-chain $\boldsymbol{\theta}_q$ with zero boundary, i.e. $\boldsymbol{\theta}_{q}\in \mbox{ker }\partial_q$ and we call a $q$-boundary a chain $\boldsymbol{\theta}_q$ that is a boundary of a higher-dimensional chain, i.e. $\boldsymbol{\theta}_{q}\in \mbox{img }\partial_{q+1}$.
The boundary operator $\partial_q$ can be encoded in the $N_{q-1}\times N_q$ boundary matrix of elements 
\be
{\mathbf{B}}_q({\mathbf{e}}_{\mu^{\prime}},{\mathbf{e}}_{\mu})=\left\{\begin{array}{cl}(-1)^p& \mbox{if }\  \mu=[i_0,i_1,\ldots, i_q] \ \mbox{and} \ \mu^{\prime}=[i_0,i_1,\ldots,i_{p-1},i_{p+1}\ldots i_q],\\0&\mbox{otherwise}\end{array}\right.
\ee
Clearly, since the boundary of the boundary is null, we have
${\mathbf{B}}_{q}{\mathbf{B}}_{q+1}=0.$
The coboundary operator $\delta_q:C^q\to C^{q+1}$ is a linear operator that maps a generic $q$-cochain $\hat{\boldsymbol{\theta}}$ into a $(q+1)$-cochain $\delta_{q}\hat{\boldsymbol{\theta}}$ of elements  
\be
\Big(\delta_{q}\hat{\boldsymbol{\theta}}\Big)_{[i_0,i_1,\ldots,i_{q+1}]}=\sum_{p=0}^{q+1}(-1)^p{\theta}_{[i_0,i_1,\ldots,i_{p-1},i_{p+1},\ldots,i_{q+1}]}.
\ee
It follows that if $\boldsymbol{\hat{\psi}}=\delta_q\hat{\boldsymbol{\theta}}\in C^{q+1}$, then, on the canonical basis, the coboundary operator can be represented by the transposed boundary matrix, as we have $\boldsymbol{\hat{\psi}}={\mathbf{B}}_{q+1}^{\top}\boldsymbol{\hat{\theta}}.$
Naturally, because the boundary of the boundary is null, we have 
$\delta_{q}\delta_{q-1}=0,$ and ${\mathbf{B}}_{q+1}^{\top}{\mathbf{B}}_{q}^{\top}=0,$
which implies $\mbox{img }({\delta}_{q-1})\subseteq \mbox{ker }({\delta}_{q})$.
We call a $q$-cocycle  a $q$-cochain $\hat{\boldsymbol{\theta}}_q$ with zero coboundary, i.e. $\hat{\boldsymbol{\theta}}_{q}\in \mbox{ker }\delta_q$ and  we call a $q$-coboundary a cochain $\hat{\boldsymbol{\theta}}_q$ that is a coboundary of a lower-dimensional cochain, i.e. $\hat{\boldsymbol{\theta}}_{q}\in \mbox{img }\delta_{q-1}$.

\subsection{The Hodge Laplacians}
The $q$-Hodge Laplacians are fundamental operators in TDL on graphs and simplicial complexes. They describe diffusion among  $q$-simplices going either through simplices of one dimension up or through simplices of one dimension down.
While allowing for independent orientation of the simplices changes some of the properties of these operators (see for details \cite{wang2026topology}) the main topological features remain unchanged. In particular the main topological properties of the graph Laplacian remain unchanged (see \cite{wang2026topology}).
In terms of the boundary matrices, the  $q$-Hodge Laplacian ${\mathbf{L}}_q$ is defined as 
\be
{\mathbf{L}}_q={\mathbf{L}}_q^{up}+{\mathbf{L}}_q^{down},\quad
\mbox{where} \quad
{\mathbf{L}}_q^{up}={\mathbf{B}}_{q+1}{\mathbf{B}}_{q+1}^{\top},\quad {\mathbf{L}}_{q}^{down}={\mathbf{B}}_{q}^{\top}{\mathbf{B}}_{q}.
\ee
Note that as a consequence of the fact that the boundary of the boundary is null we have  that the up and down Laplacians do not only commute, but their product is also zero, i.e.
${\mathbf{L}}_q^{up}{\mathbf{L}}_q^{down}=0,$ and ${\mathbf{L}}_q^{down}{\mathbf{L}}_q^{up}=0.$
Specifically, in the case of a graph ($1$-dimensional simplicial complex) the relevant boundary operator is exclusively ${\mathbf{B}}_1$ while the Hodge Laplacians include the graph Laplacian ${\mathbf{L}}_0={\mathbf{B}}_1{\mathbf{B}}_1^{\top}$ and the $1$-Laplacian ${\mathbf{L}}_{1}={\mathbf{B}}_1^{\top}{\mathbf{B}}_1$.
Interestingly, starting from the boundary operator it is also possible to construct Dirac operators defined as the ``square root" of the Laplacians \citep{bianconi2021topological}.
\section{Existing Simplicial Complex Neural Networks}
 The Hodge Laplacians  have a fundamental role in topological dynamics \citep{millan2025topology,millan2020explosive} and  signal processing \citep{barbarossa2020topological} which have significantly inspired the formulation of topological neural network architectures. \cite{ebli2020simplicialneuralnetworks} propose a convolutional neural network that uses the boundary operators to define simplicial Laplacians for each simplicial convolutional layer. \cite{roddenberry2021} and \cite{yang2022} extend this by treating upper and lower boundary operators separately in each convolutional layer, while \citep{yang2023} further incorporate  propagation between simplices of the same dimension. The Message Passing Simplicial Network \citep{bodnar2021} introduces four types of message passing  all of which can be defined using boundary matrices. \cite{10735150} introduce self-attention layers relying on the Dirac operator \citep{bianconi2021topological}, previously successfully adopted for signal processing \citep{calmon2023dirac,wang2025dirac}, to define the convolution. \cite{goh2022} introduce another simplicial attention neural network and, while the architecture does not rely on boundary matrices, only the propagation between simplices of adjacent dimensions is considered. Similarly, \cite{TANG2026} propose the Hyperbolic simplicial convolutional network which maps network embeddings in hyperbolic space and considers upper and lower neighbourhoods for convolution. Other proposed architectures  including concepts such as locality sensitive hashing \citep{tan2023}, block Hodge Laplacians \citep{chen2021}, conditioning on $E(n)$ equivariant information \citep{pmlr-v202-eijkelboom23a} and Quantum Neural Networks \citep{Piperno:2025nlx} still rely on boundary operators for their convolutional steps or message-passing. More recently, \cite{3780338.3782254} proposed TopoTune, a framework that combines generic boundary-based message passing approaches to obtain a general and flexible architecture.
These proposed network architectures either rely on the boundary operators or consider only adjacencies, limiting the expressive power of the models.

\section{Research question of this work }
In this work, we formulate CCNN to predict node features by taking into account the information encoded in the higher-order simplices, considering not only edges but also triangles, etc. For instance, in social networks there is a well-established assumption \citep{iacopini2019simplicial} that the presence of triangles and cliques strongly biases the opinion dynamics occurring at the node level.
Inspired by these considerations, we aim to design the CCNN architecture that is suitable for tasks involving this type of dynamics.
Interestingly, this research question cannot be tackled by traditional TNN approaches based on the boundary operators and the Hodge Laplacians. Indeed, the same fundamental topological property, namely that the boundary of the boundary is null, implies that if we only rely on the boundary operators, topological signals associated with the nodes cannot be lifted to triangles or to higher-dimensional simplices in a single layer. Therefore, their propagation cannot take into account the higher-order topological structure of the data.
In order to address our research question, we therefore need to consider alternative mathematical frameworks that go beyond the exclusive use of the boundary operators and the Hodge Laplacians. In particular, in this work we formulate the Cup and Cap Topological Neural Network (CCNN), based on the cup and cap products.  As we will show, these products provide a natural framework for lifting and lowering cochains. In this way, the CCNN incorporates the higher-order topological structure of the simplicial complex into the prediction of node features.

\section{The benefits of the cup and cap products}
In this section we summarise the definition of the cup and cap products \citep{nanda2021computational,hatcher2005algebraic,munkres2018elements} which are key for formulating the CCNN and allow us to  lift  and lower the topological signals across simplices of arbitrary dimension in a single layer.

\subsection{Cup product as a lifting operator}
Consider the $p$-cochain $\hat{\boldsymbol{\theta}}\in C^p$ and the $q$-cochain $\hat{\boldsymbol{\phi}}\in C^q$. The cup product $\hat{\boldsymbol{\theta}}\smile \hat{\boldsymbol{\phi}}\in C^{p+q}$ is the $(p+q)$-cochain that has elements 
\be
(\hat{\boldsymbol{\theta}}\smile\hat{\boldsymbol{\phi}})=\sum_{(i_0,i_1,\ldots,i_{p+q})\in Q_{p+q}}\theta_{[i_0,i_1,\ldots,i_p]} \phi_{[i_p,i_{p+1},\ldots, i_{p+q}]}{\hat{\mathbf{e}}}_{[i_0,i_1,\ldots,i_{p+q}]}
\ee
Thus the cup product can be used to lift cochains.
Examples of cup products include the following: 
\begin{itemize}
\item[(1)]
A  1-cochain $\hat{\boldsymbol{\xi}}\in C^1$ can be built exclusively from a 0-cochain $\hat{\boldsymbol{\theta}}\in C^0$ and the $1$-cochain $\hat{\mathbf{1}}_{N_{1}}\in C^1$ which   has elements equal to one on every edge of the network, i.e.
\be
\hat{\boldsymbol{\xi}}=\hat{\boldsymbol{\theta}}\smile \hat{\mathbf{1}}_{N_{1}}\smile\hat{\boldsymbol{\theta}}\quad \mbox{with elements} \quad {\xi}_{[rs]}=\theta_{[r]}\theta_{[s]}.
\ee
This procedure is used in \cite{faskowitz2020edge} to produce a 1-cochain from a 0-cochain in order to define edge signals in the brain (see also \cite{santoro2023higher}), although the relation to the cup product is not emphasised.
As we will see, we will use this cup product to lift $0$-cochains in the CCNN architecture.
\item[(2)] A 1-cochain $\hat{\boldsymbol{\xi}}\in C^1$ can be  built by combining a 0-cochain $\hat{\boldsymbol{\theta}}\in C^0$ with a 1-cochain $\hat{\boldsymbol{\phi}}\in C^1$ as in 
\be
\hat{\boldsymbol{\xi}}=\hat{\boldsymbol{\theta}}\smile \hat{\boldsymbol{\phi}}\quad
\mbox{with elements}\quad
{\xi}_{[rs]}=\theta_{[r]}\phi_{[rs]}
\ee
\item[(3)]
A 2-cochain $\hat{\boldsymbol{\psi}}\in C^2$ can be built from two 1-cochains $\hat{\boldsymbol{\phi}}\in C^1$ and $\hat{\boldsymbol{\xi}}\in C^1$  as 
\be
\hat{\boldsymbol{\psi}}=\hat{\boldsymbol{\phi}}\smile \hat{\boldsymbol{\xi}}\quad\mbox{with elements}\quad
{\psi}_{[rsp]}=\phi_{[rs]}\xi_{[sp]}.
\ee
An example of this is an operation that we will use in the architecture of the CCNN: 
$\hat{\boldsymbol{\psi}}=\hat{\boldsymbol{\xi}}\smile {\mathbf{B}}_1^{\top}\hat{\boldsymbol{\theta}}$.
Note that $\hat{\boldsymbol{\xi}}$ in this example can  be a $1$-cochain derived exclusively from a $0$-cochain either by using example (1) or by using an element-wise even function of ${\mathbf{B}}_1^{\top}\hat{\boldsymbol{\theta}}$, i.e. $\hat{\boldsymbol{\xi}}=g({\mathbf{B}}_1^{\top}\hat{\boldsymbol{\theta}})$.  It follows that  in this way one can lift a 0-cochain to a 2-cochain, overcoming the limitation due to the exclusive use of the boundary operators and the Hodge Laplacians.
\end{itemize}

Interestingly, it can be shown \citep{hatcher2005algebraic}  that  the cup product of two cocycles is a cocycle, and the cup product of a cocycle and a coboundary is a coboundary.

\subsection{Cap product as a lowering operator}
Consider the $p$-cochain $\hat{\boldsymbol{\theta}}\in C^p$  and the $q$-chain $\boldsymbol{\phi}\in C_q$  with $q\geq p$ given by
\be
\hat{\boldsymbol{\theta}}=\sum_{(i_0,i_1,i_2,\ldots,i_p)\in Q_p}\theta_{[i_0,i_1,i_2,\ldots,i_p]}\hat{\mathbf{e}}_{[i_0,i_1,\ldots,i_p]},\quad
\boldsymbol{\phi} = \sum_{(i_0,i_1,i_2,\ldots,i_q)\in Q_q}\phi_{[i_0,i_1,i_2,\ldots,i_q]} {{\mathbf{e}}}_{[i_0,i_1,\ldots,i_q]}.
\ee 
The cap product $ \hat{\boldsymbol{\theta}}\frown{\boldsymbol{\phi}}\in C_{q-p}$ is the $(q-p)$-chain of elements
\be
\hat{\boldsymbol{\theta}}\frown{\boldsymbol{\phi}}=\sum_{(i_0,i_1,i_2,\ldots,i_q)\in \mathcal{Q}_q}\theta_{[i_0,i_1,\ldots i_p]}\phi_{[i_0,i_1,\ldots, i_q]}{\mathbf{e}}_{[i_p,i_{p+1},\ldots, i_{q}]}
\ee
Thus, the cap product can be used as a lowering operator that transforms cochains into lower-dimensional chains.
Examples of the cap product include:
\begin{itemize}

\item[(1)]
Consider a 1-cochain  $\hat{\boldsymbol{\xi}}\in C^1$ and the 1-chain $\boldsymbol{\alpha}^{(1)}\in C_1$ constructed from the adjacency matrix ${\mathbf{a}}^{(1)}$ given by 
\be
\boldsymbol{\alpha}^{(1)}=\sum_{i_0,i_1}a^{(1)}_{i_0 i_1}{\mathbf{e}}_{[i_0,i_1]}.
\ee
We can construct the 0-chain $\boldsymbol{\eta}\in C_0$ as 
\be
\boldsymbol{\eta}=\hat{\boldsymbol{\xi}}\frown \boldsymbol{\alpha}^{(1)}\quad \mbox{with elements}
\quad 
\eta_{[r]}=\sum_{s=1}^N\xi_{[sr]}a^{(1)}_{sr}.
\ee
Interestingly, if $\hat{\boldsymbol{\xi}}$ is the coboundary of a $0$-cochain $\hat{\boldsymbol{\theta}}\in C^0$, i.e. if $\hat{\boldsymbol{\xi}}={\mathbf{B}}_1^{\top}\hat{\boldsymbol{\theta}}$, we obtain 
\be
\eta_{[r]}=\sum_{s=1}^N(\theta_{[r]}-\theta_{[s]})a^{(1)}_{[sr]}=\sum_s[{\hat{\mathbf{L}}}_{[0]}]_{rs}\theta_{[s]}=\Big({\hat{\mathbf{L}}}_{[0]}\boldsymbol{\theta}\Big)_{[r]}.
\ee
where here we indicate with ${\hat{\mathbf{L}}}_{[0]}$ is the standard graph-Laplacian constructed by a $1$-boundary operator where edges have a single positive orientation.
\item[(2)]
Consider now the  2-cochain $\hat{\boldsymbol{\psi}}\in C^2$ and the  2-chain $\boldsymbol{\alpha}^{(2)}\in C_2$ constructed from the adjacency tensor ${\mathbf{a}}^{(2)}$ given by 
\be
\boldsymbol{\alpha}^{(2)}=\sum_{i_0=1}^N\sum_{i_1=1}^N\sum_{i_2=1}^Na^{(2)}_{i_0 i_1 i_2}{\mathbf{e}}_{[i_0,i_1,i_2]}.
\ee
We can construct the $0$-chain $\boldsymbol{\eta}\in C_0$ as 
\be
\boldsymbol{\eta}=\hat{\boldsymbol{\psi}}\frown \boldsymbol{\alpha}^{(2)}\quad \mbox{with}
\quad \eta_{[r]}=\sum_{m=1}^N\sum_{s=1}^N\psi_{[msr]}a^{(2)}_{[msr]}.
\ee
\end{itemize}

Interestingly, it can be shown \citep{hatcher2005algebraic} that the cap product between a cocycle and a cycle is a cycle while the cap product between a coboundary and a cycle is a boundary.

\section{Cup and Cap Topological Neural Network (CCNN) architecture}
The CCNN is a higher-order topological neural network that builds on the mathematical properties of the cup  and cap products to learn node features by lifting the 0-cochain to higher-dimensional cochains and then lowering the signal to nodes again. In this way, the CCNN is designed to use the information about the higher-dimensional topological structure of the data to solve node-based tasks.
The CCNN architecture consists of $L$ layers  $\ell\in \{1,\ldots, L\}$, each with hidden dimension $F^{\ell}$. We indicate with $M$ the largest dimension of lifted signals.
As mentioned before, the CCNN architecture takes as input a 0-cochain $\hat{\boldsymbol{\theta}}^0$ where we extend the definition by assuming that its elements are defined in $\mathbb{R}^n$, i.e. $\hat{\boldsymbol{\theta}}^0\in \mathbb{R}^{N_0\times n}$ where $n$ indicates the dimension of the input features associated with the nodes. This extension is equivalent to treating $\hat{\boldsymbol{\theta}}^0$ as the direct sum of $n$ 0-cochains.
While the input does not include an edge signal, the intermediate layers of the CCNN are topological and include the $0$-cochain $\hat{\boldsymbol{\theta}}^{\ell}\in \mathbb{R}^{N_0\times F^{\ell}}$ together with the internally defined $q$-cochains $\hat{\boldsymbol{\chi}}^{\ell}_q\in \mathbb{R}^{N_q\times F^{\ell}}$.

The learnable terms include two sets of matrices ${\mathbf{W}}_q^{\ell}\in \mathbb{R}^{F^{\ell-1}\times F^{\ell}}$ and $\hat{\mathbf{W}}_q^{\ell}\in \mathbb{R}^{F^{\ell-1}\times F^{\ell}}$   and a set of learnable  matrices ${\mathbf{D}}^{\ell}\in \mathbb{R}^{F^{\ell-1}\times F^{\ell}}$, $\hat{\mathbf{D}}^{\ell}_q\in \mathbb{R}^{F^{\ell-1}\times F^{\ell}}$  and $\tilde{\mathbf{D}}^{\ell}_q\in \mathbb{R}^{N_q\times F^{\ell}}$ with $q\in \{1,2,\ldots,M\}$ and $\ell\in \{1,\ldots, L\}$. 

The CCNN is defined through a recursive system of equations for the cochains $\hat{\boldsymbol\theta}^{\ell}\in C^0$ (and their associated chains $\boldsymbol\theta^{\ell}\in C_0$)  and the auxiliary $q$-cochains $\hat{\boldsymbol\chi}_q^{\ell}\in C^q$. The recursive equation for  $\hat{\boldsymbol\theta}^{\ell}\in C^0$ and ${\boldsymbol\theta}^{\ell}\in C_0$ is given by 
\begin{align}
{\boldsymbol{\theta}}^{\ell+1}=\sigma\left({\hat{\boldsymbol{\theta}}}^{\ell}{\mathbf{D}}^{\ell+1}-\sum_{q=1}^M \Big(f({\hat{\boldsymbol{\chi}}}^{\ell}_{q-1})\smile({\mathbf{B}}_1^{\top}{\hat{\boldsymbol{\theta}}}^{\ell})\Big){\mathbf{W}}^{\ell+1}_{q}\frown {\boldsymbol{\alpha}}^{(q)}\right).
\end{align}
Here $\sigma(x)$ indicates the non-linear function of choice and $f(x)$ is taken to be  $f(x)=\tanh(x)$ if $\sigma$ is  ReLU or $f(x)=x$  if $\sigma$ is the sigmoid function. 
In this work  $\sigma$ is always taken to be the ReLU. Note that here and in the following we always assume that $f({\hat{\boldsymbol{\chi}}}_q^{\ell})$ does not depend on the orientation of the simplices. Specifically, indicating with $f(\chi^{\ell}_{[i_0,i_1,\dots i_q]})$ the generic element of $f(\hat{\boldsymbol{\chi}}_q^{\ell})$ we always assume $f(\chi^{\ell}_{[i_0,i_1,\dots i_q]})=f(\chi^{\ell}_{[\pi(0),\pi(1),\ldots \pi(q)]}),$ where $\pi$ is an arbitrary permutation of the indices.
Therefore, for the purposes of the CCNNs, it is sufficient to specify $\hat{\boldsymbol{\chi}}_q^{\ell}$ for a single positive orientation of the simplices.

We propose two different Cup and Cap Topological Neural Networks (CCNNs) selected after an ablation study of different variants of the architecture (see Appendix \ref{apd:ablation}). The CCNN models differ by the choice of the recursive equation for the auxiliary $q$-cochain $\hat{\boldsymbol\chi}^{\ell}_q\in C^q$ built from the 0-cochain $\hat{\boldsymbol{\theta}}^{\ell}\in C^0$ using different cup and cap liftings.  
The first CCNN-$\eta$ version is characterized by the following recursive equations for $\hat{\boldsymbol\chi}^{\ell}_q$ with $q>0$ 
\begin{align}
\hat{\boldsymbol{\chi}}^{\ell+1}_q=\sigma\Big(\hat{\boldsymbol{\chi}}^{\ell}_{q}{\hat{\mathbf{D}}_q^{\ell+1}}+\Big(\hat{\boldsymbol{\chi}}^{\ell}_{q-1}\smile\left|{\mathbf{B}}_1^{\top}\hat{\boldsymbol{\theta}}^{\ell}\right|\Big)\hat{\mathbf{W}}^{\ell+1}_{q}+\tilde{\mathbf{D}}_q^{\ell+1}\Big),
\label{eq:chi1}
\end{align}
 while for $q=0$ we have $\hat{\boldsymbol{\chi}}^{\ell}_0={\mathbf{1}}.$ 
 The CCNN-$\omega$ version instead adopts the following recursive equations for $\hat{\boldsymbol\chi}^{\ell}_q$ with $q>0$ 
\begin{align}
\hat{\boldsymbol{\chi}}_q^{\ell+1}=\sigma(\hat{\boldsymbol{\chi}}_q^{\ell}\hat{\mathbf{D}}_{q}^{\ell+1}+\Big(\hat{\boldsymbol{\chi}}_{q-1}^{\ell}\smile\hat{\boldsymbol{\theta}}^{\ell}\smile{\hat{\mathbf{1}}}_{N_1}\smile \hat{\boldsymbol{\theta}}^{\ell}\Big){\mathbf{W}}_{q}^{\ell+1}+\tilde{\mathbf{D}}_{q}^{\ell+1}).
\label{eq:chi2}
\end{align}

In this work we focus specifically on CCNNs with dimension $M=2$ defined on simplicial complexes formed by nodes, edges and triangles, where the lifted node signal propagates through simplices of dimension up to $M=2$ including edges and triangles. Therefore, in this case, the only required auxiliary cochain is the $1$-cochain ${\hat{\boldsymbol{\chi}}}_1$ which we will indicate simply by $\hat{\boldsymbol{\chi}}$. 
In this scenario, by identifying $C_0$ and $C^0$ through the canonical basis, the  $M=2$ CCNNs  can also be expressed in matrix form as a set of recursive  equations for the cochains $\hat{\boldsymbol{\theta}}^{\ell}\in \mathbb{R}^{N_0\times F_{\ell}}, \hat{\boldsymbol{\chi}}^{\ell}\in \mathbb{R}^{N_1\times F_{\ell}}$. Finally, we relabel the learnable matrices using indices  $(j)$ with $1\leq j\leq 6$. Thus,  the CCNN recursive equations for  $\hat{{\boldsymbol{\theta}}}^{\ell+1}$ read 
\be
\hat{{\boldsymbol{\theta}}}^{\ell+1}=\sigma\left(\hat{\boldsymbol{\theta}}^{\ell}{{\mathbf{D}}^{\ell+1}}_{(1)}+\hat{\mathbf{L}}_{0}\hat{\boldsymbol{\theta}}^{\ell}{\mathbf{W}}^{\ell+1}_{(2)}+{\mathbf{T}}(\hat{\boldsymbol{\theta}}^{\ell},\hat{\boldsymbol{\chi}}^{\ell}){\mathbf{W}}_{(3)}^{\ell+1}\right), 
\ee where $\hat{\bf L}_{[0]}$ is the graph Laplacian corresponding to a single positive orientation of the simplices, and  ${\mathbf{T}}(\hat{\boldsymbol{\theta}}^{\ell}, \hat{\boldsymbol{\chi}}^{\ell})\in \mathbb{R}^{N_0\times F_{\ell}}$ has elements
\be
[{\mathbf{T}}(\hat{\boldsymbol{\theta}}^{\ell},\hat{\boldsymbol{\chi}}^{\ell})]_{[r]}=\sum_{s<m}a_{rsm}f(\hat{\boldsymbol{\chi}}^{\ell}_{[sm]})(\hat{\boldsymbol{\theta}}^{\ell}_{[s]}+\hat{\boldsymbol{\theta}}^{\ell}_{[m]}-2 \hat{\boldsymbol{\theta}}^{\ell}_{[r]}).
\ee
The $1$-cochain $\hat{\boldsymbol{\chi}}^{\ell}$ is given for the CCNN-$\eta$ as 
\be
\hat{\boldsymbol{\chi}}^{\ell+1}=\sigma\Big(\hat{\boldsymbol{\chi}}^{\ell}{\mathbf{D}}_{(4)}^{\ell+1}+\left|{\mathbf{B}}^{\top}_{1}\hat{\boldsymbol{\theta}}^{\ell}\right|{\mathbf{W}}_{(5)}^{\ell+1}+{\mathbf{D}}_{(6)}^{\ell+1}\Big),\ee
while  for the CCNN-$\omega$ it is given by 
\be
\hat{\boldsymbol{\chi}}^{\ell+1}=\sigma(\hat{\boldsymbol{\chi}}^{\ell}{\mathbf{D}}_{{(4)}}^{\ell+1}+\Big(\hat{\boldsymbol{\theta}}^{\ell}\smile{\hat{\mathbf{1}}}_{N_1}\smile \hat{\boldsymbol{\theta}}^{\ell}\Big){\mathbf{W}}_{(5)}^{\ell+1}+{\mathbf{D}}_{(6)}^{\ell+1}),
\label{eq:chi2_M2}
\ee
where 
$\hat{\boldsymbol{\chi}}^{\ell=0}={\mathbf{B}}^{\top}_{1}\hat{\boldsymbol{\theta}}^{\ell=0}.$

\section{Experimentation}
We evaluate our novel architecture on TopoBench \citep{telyatnikov2025}, a library designed to benchmark Topological Deep Learning architectures. This library contains  a variety of predefined liftings through which  a set of graph datasets can be transformed into hypergraphs or simplicial complexes. To enable a fair comparison between our CCNN architecture and the previously proposed simplicial complex architectures, we apply the Clique Lifting operator to the graph datasets \citep{telyatnikov2025}, i.e., any graph is lifted to a simplicial complex by treating each $n$-clique as an $(n-1)$-simplex. We also consider a fixed hidden dimension $F^{\ell} = F^{\ell-1}$ and ${\bf D}_{(1)}^{\ell+1},{\bf D}_{(4)}^{\ell+1}$  to be diagonal matrices, and ${\bf D}_{(6)}^{\ell+1}$ to be a bias, for parameter efficiency. To transform the input data into features of the right dimension, and the hidden representation into the final prediction, we use the same encoder and readout used by \cite{telyatnikov2025}.

\begin{table}[tbp!]
\centering
\caption{Cross-domain comparison: results are shown as mean and standard deviation. The best result is shown in bold and shaded in grey, while those within one standard deviation are in blue-shaded boxes. ``TopoBench (best)'' reports the best-performing baseline specified by \citet{telyatnikov2025} that is not the simplicial complex domain.}
\label{table:main}
\begin{adjustbox}{width=1.\textwidth}
  \label{tab:main-results}
  \begin{tabular}{@{}llrr|rrr|r@{}}
\toprule
&
\multicolumn{1}{l}{Dataset} &
\multicolumn{1}{c}{CCNN-$\eta$} &
\multicolumn{1}{c}{CCNN-$\omega$} &
\multicolumn{1}{|c}{SCN} &
\multicolumn{1}{c}{SCCN} &
\multicolumn{1}{c}{SCCNN} &
\multicolumn{1}{|c}{TopoBench (best)} \\
\midrule
\multirow{14}{*}{\rotatebox[origin=c]{90}{Node-level tasks}}
& Cora & 87.68 ± 1.14 & 87.39 ± 0.79 & 82.27 ± 1.34 & 80.86 ± 2.16 & 82.19 ± 1.07 & \cellcolor{gray!25} \textbf{88.92 ± 0.44} {\scriptsize(AST)} \\

 & Citeseer & 73.52 ± 1.61 & 73.61 ± 1.42 & 71.24 ± 1.68 & 69.60 ± 1.83 & 70.23 ± 2.69 & \cellcolor{gray!25} \textbf{75.63 ± 1.58} {\scriptsize(CCCN)} \\

 & Pubmed & \cellcolor{gray!25} \textbf{89.74 ± 0.31} & \cellcolor{blue!25} 89.68 ± 0.66 & 88.72 ± 0.50 & 88.37 ± 0.48 & 88.18 ± 0.32 & \cellcolor{blue!25} 89.62 ± 0.25 {\scriptsize(AST)} \\

 & Amazon & \cellcolor{gray!25} \textbf{52.22 ± 0.81} & 50.99 ± 0.97 & OOM & OOM & OOM & \cellcolor{blue!25} 51.90 ± 0.15 {\scriptsize(CWN)} \\

 & Empire & 84.46 ± 0.67 & 83.85 ± 0.55 & 88.79 ± 0.46 & 88.27 ± 0.14 & \cellcolor{gray!25} \textbf{89.15 ± 0.32} & 84.02 ± 0.51 {\scriptsize(GAT)} \\

 & Minesweeper & 87.22 ± 0.56 & 87.76 ± 0.31 & \cellcolor{gray!25} \textbf{90.32 ± 0.11} & 89.07 ± 0.25 & 89.0 ± 0.00 & 89.42 ± 0.00 {\scriptsize(CCCN)} \\

 & Election & \cellcolor{blue!25} 0.29 ± 0.01 & 0.31 ± 0.01 & 0.46 ± 0.04 & 0.53 ± 0.03 & 0.51 ± 0.03 & \cellcolor{gray!25} \textbf{0.28 ± 0.02} {\scriptsize(GIN)} \\

 & Bachelor & \cellcolor{blue!25} 0.30 ± 0.03 & 0.33 ± 0.04 & 0.32 ± 0.02 & 0.36 ± 0.02 & 0.34 ± 0.03 & \cellcolor{gray!25} \textbf{0.28 ± 0.02} {\scriptsize(GAT)} \\

 & Birth & \cellcolor{blue!25} 0.74 ± 0.10 & \cellcolor{blue!25} 0.71 ± 0.08 & \cellcolor{blue!25} 0.71 ± 0.08 & 0.82 ± 0.09 & 0.79 ± 0.12 & \cellcolor{gray!25} \textbf{0.70 ± 0.07} {\scriptsize(EDGNN)} \\

 & Death & \cellcolor{blue!25} 0.51 ± 0.04 & \cellcolor{blue!25} 0.52 ± 0.04 & \cellcolor{blue!25} 0.52 ± 0.05 & 0.58 ± 0.06 & 0.55 ± 0.05 & \cellcolor{gray!25} \textbf{0.49 ± 0.05} {\scriptsize(AST)} \\

 & Income & \cellcolor{blue!25} 0.22 ± 0.02 & 0.23 ± 0.02 & 0.25 ± 0.02 & 0.29 ± 0.03 & 0.28 ± 0.03 & \cellcolor{gray!25} \textbf{0.20 ± 0.02} {\scriptsize(GAT)} \\

 & Migration & \cellcolor{blue!25} 0.84 ± 0.13 & \cellcolor{blue!25} 0.85 ± 0.16 & 0.92 ± 0.20 & 0.91 ± 0.18 & \cellcolor{blue!25} 0.90 ± 0.14 & \cellcolor{gray!25} \textbf{0.77 ± 0.13} {\scriptsize(GAT)} \\

 & Unempl & \cellcolor{blue!25} 0.23 ± 0.02 & 0.26 ± 0.03 & 0.38 ± 0.04 & 0.43 ± 0.04 & 0.43 ± 0.04 & \cellcolor{gray!25} \textbf{0.22 ± 0.02} {\scriptsize(GIN/AST)} \\
\midrule
\multirow{8}{*}{\rotatebox[origin=c]{90}{Graph-level tasks}}
& MUTAG & \cellcolor{gray!25} \textbf{82.55 ± 5.90} & \cellcolor{gray!25} \textbf{82.55 ± 5.28} & 73.62 ± 6.13 & 70.64 ± 5.9 & 76.17 ± 6.63 & 80.43 ± 1.78 {\scriptsize(CWN)} \\

 & PROTEINS & 74.70 ± 0.92 & 74.41 ± 2.59 & \cellcolor{blue!25} 75.27 ± 2.14 & \cellcolor{blue!25} 75.05 ± 2.76 & 74.19 ± 2.86 & \cellcolor{gray!25} \textbf{76.63 ± 1.74} {\scriptsize(AST)} \\

 & NCI1 & \cellcolor{gray!25} \textbf{76.89 ± 0.98} & \cellcolor{blue!25} 76.19 ± 0.83 & 74.49 ± 1.03 & \cellcolor{blue!25} 76.17 ± 1.39 & \cellcolor{blue!25} 76.60 ± 1.75 & \cellcolor{blue!25} 76.67 ± 1.48 {\scriptsize(CCCN)} \\

 & NCI109 & 75.61 ± 1.04 & \cellcolor{blue!25} 76.55 ± 1.13 & 75.70 ± 1.04 & 75.49 ± 1.39 & \cellcolor{gray!25} \textbf{77.12 ± 1.07} & 75.66 ± 1.30 {\scriptsize(CCXN)} \\

 & IMDB-BIN & \cellcolor{blue!25} 70.24 ± 1.44 & \cellcolor{blue!25} 69.84 ± 2.91 & \cellcolor{blue!25} 70.80 ± 2.38 & \cellcolor{blue!25} 70.88 ± 3.98  & \cellcolor{blue!25} 70.88 ± 2.25 & \cellcolor{gray!25} \textbf{72.00 ± 2.48} {\scriptsize(GCN)} \\

 & IMDB-MUL & \cellcolor{blue!25} 48.75 ± 3.8 & \cellcolor{blue!25} 48.11 ± 3.67 & \cellcolor{blue!25} 49.49 ± 5.08 &\cellcolor{blue!25} 49.71 ± 3.7  & \cellcolor{blue!25} 48.75 ± 3.98 & \cellcolor{gray!25} \textbf{50.51 ± 2.92} {\scriptsize(AST)} \\

 & ZINC & 0.55 ± 0.01 & 0.54 ± 0.01 & 0.53 ± 0.04 &0.46 ± 0.08 & 0.36 ± 0.02 & \cellcolor{gray!25} \textbf{0.34 ± 0.01} {\scriptsize(CWN/CCCN)} \\
\bottomrule
  \end{tabular}
\end{adjustbox}
\end{table}

We benchmark CCNN-$\eta$ and CCNN-$\omega$ against previously proposed simplicial neural networks: SCN \citep{ebli2020simplicialneuralnetworks}, SCCN \citep{yang2022EfficientRepresentation} and SCCNN \citep{yang2023}, as well as the strongest baseline across all other higher-order topological domains reported by \cite{telyatnikov2025}. Table \ref{tab:main-results} shows the results on regression and classification tasks averaged over 5 random seeds. CCNN-$\eta$ and CCNN-$\omega$ outperform the other simplicial complex architectures on $8$ out of $20$ tested datasets. On $13$  datasets, their performance is within one standard deviation of the best model tested in TopoBench \citep{telyatnikov2025} across all domains including graph, hypergraph, simplicial complex and combinatorial complex. In fact, CCNN-$\eta$ beats all other topological baselines on Pubmed, Amazon, MUTAG and NCI1 and CCNN-$\omega$ on MUTAG. This provides evidence of the advantages of using the cup and cap products to leverage the topological structure of the data. 

\section{Conclusions}

Despite their success, TDL architectures based on the boundary operators and the Hodge Laplacians have the intrinsic limitation that they cannot lift or lower signals across more than one dimension per layer. Here we demonstrate that TDL architectures based on the cup and cap products can overcome this limitation.
Specifically, we propose the CCNN, a TDL architecture that enables node-based predictions by exploiting the higher-order interactions present in the data, such as triangles. 
The CCNN shows competitive performance against models from different domains and either matches or improves performance with respect to previous simplicial-complex-based neural networks for the majority of the tested datasets, establishing the CCNN as a valid option for prediction tasks. The CCNN architecture only relies on node features and can easily be applied to a wide range of datasets and prediction tasks. Overall, our work showcases the potential of using the cup and cap products for Topological Deep Learning. This provides an alternative to the standard Hodge-Laplacian-based approach, thus opening up new research directions for exploring TDL architectures based on the cup and cap products.

 \section*{Author Contributions}
G.B. designed the project, M.N. and F.H.C.  led the methodological development and execution of the project. All authors contributed through discussions, feedback, interpretation of the results and writing of the manuscript.

 \acks{G.B. acknowledges interesting discussions with V. Nanda, S. Barbarossa and L. Di Nino. M.N. was funded by the UKRI/BBSRC Collaborative Training Partnership in AI for Drug Discovery and Queen Mary University of London. This research utilised Queen Mary's Apocrita HPC facility, supported by QMUL Research-IT. http://doi.org/10.5281/zenodo.438045. F. H. C. acknowledges  support of a fellowship from “la Caixa” Foundation (ID 100010434). The fellowship code is LCF/BQ/PFA25/11000012.
 }

\bibliography{pmlr-sample}

\appendix

\section{Hyperparameter sweeps }\label{apd:exp_setup}

In our experiments we  use the All Cell Feature Encoder as our encoder. We have designed a  custom CCNN backbone and  a custom CCNN wrapper that encodes the Cup and Cap Topological Neural Network. As a readout we consider two options: Signal Down Propagation and no specific readout. This will be treated as one of our hyperparameters. We consider the same hyperparameters as in TopoBench, which further include learning rate $(0.01,0.001)$, hidden dimensions $(32,64,128)$, the number of layers of the forward pass $(2,3,4,5)$ and the encoder dropout $(0.25,0.5)$. Additionally, when we compare the performance of CCNN against other GNNs and TNNs, we use normalisation and dropout techniques in our backbone. We also apply early stopping in our experiments.
\section{Ablation study}\label{apd:ablation}
To choose the best architecture for our CCNN backbone, we use an ablation study. The different options we consider are depicted in Table \ref{tab:options}. $A$, $B$ and $C$ are the individual terms of our forward pass, which rely on information from the nodes, edges and  triangles, respectively. Options $\gamma$ to $\omega$ show different ways of calculating the auxiliary cochain $\hat\chi$. For our ablation study we consider two datasets, Cora and Citeseer \citep{sen:aimag08}. 
 We compare 22 different architectures in Tables  \ref{tab:ablation_Citeseer} and \ref{tab:ablation_Cora}. There is no architecture that outperforms all other architectures across both datasets. We choose architecture $A + B + C$. This allows for the greatest flexibility as the neural network can learn which of the A, B and C terms are relevant for the specific task and adjust weights accordingly. We choose the $\eta$ and the $\omega$ versions as our options for $\hat\chi$ as they perform generally well for the Citeseer data and they also yield performance within one standard deviation of the best among the $A+B+C$ options for Cora.
\begin{table*}
    \centering
     \hspace*{-10mm}\begin{tabular}{||c|c||}
     \hline\hline
     $A$&$\hat{\boldsymbol{\theta}^{\ell}}{{\mathbf{D}}^{\ell+1}}_{(1)}$\\
     $B$& ${\mathbf{L}}_{0}\hat{\boldsymbol{\theta}}^{\ell}{\mathbf{W}}^{\ell+1}_{(2)}$\\
     $C$ & ${\mathbf{T}}(\hat{\boldsymbol{\theta}}^{\ell},\hat{\boldsymbol{\chi}}^{\ell}){\mathbf{W}}_{(3)}^{\ell+1}$\\
     \hline
     $\gamma$&$\hat{\boldsymbol{\chi}}^{\ell+1}=\sigma(\hat{\boldsymbol{\chi}}^{\ell}{\mathbf{D}}^{\ell+1}_{{(4)}}+|{\mathbf{B}}^{\top}_{1}\hat{\boldsymbol{\theta}}^{\ell}|{\mathbf{W}}_{(5)}^{\ell+1})$\\
    $\delta$ &$\hat{\boldsymbol{\chi}}^{\ell+1}=\sigma(\hat{\boldsymbol{\chi}}^{\ell}{\mathbf{D}}_{{(4)}}+|{\mathbf{B}}^{\top}_{1}\hat{\boldsymbol{\theta}}^{\ell}|^2{\mathbf{W}}_{(5)}^{\ell+1})$\\
    $\eta$&$\hat{\boldsymbol{\chi}}^{\ell+1}=\sigma(\hat{\boldsymbol{\chi}}^{\ell}{\mathbf{D}}^{\ell+1}_{{(4)}}+|{\mathbf{B}}^{\top}_{1}\hat{\boldsymbol{\theta}}^{\ell}|{\mathbf{W}}_{(5)}^{\ell+1}+{\mathbf{D}}^{\ell+1}_{(6)})$\\
    $\zeta$&$\hat{\boldsymbol{\chi}}^{\ell+1}=\sigma(\hat{\boldsymbol{\chi}}^{\ell}{\mathbf{D}}^{\ell+1}_{{(4)}}+|{\mathbf{B}}^{\top}_{1}\hat{\boldsymbol{\theta}}^{\ell}|^2{\mathbf{W}}_{(5)}^{\ell+1}+{\mathbf{D}}^{\ell+1}_{(6)})$\\
    $\omega$&$\hat{\boldsymbol{\chi}}^{\ell+1}=\sigma(\hat{\boldsymbol{\chi}}^{\ell}{\mathbf{D}}^{\ell+1}_{{(4)}}+\Big(\hat{\boldsymbol{\theta}}^{\ell}\smile {\mathbf{1}}\smile \hat{\boldsymbol{\theta}}^{\ell}\Big){\mathbf{W}}_{(5)}^{\ell+1}+{\mathbf{D}}^{\ell+1}_{(6)})$\\
    \hline\hline
    \end{tabular}
    \caption{Different architectural options for the CCNN backbone architecture.}
    \label{tab:options}
\end{table*}
\begin{table*}[]
    \centering
    \hspace*{-10mm}\begin{tabular}{||c|cccc|cc||}
    \hline\hline
         & $A+B+C$ & $B+C$ & $C$ &$A+C$ & $B$ & $A+B$\\
        $\gamma$
        & \colorbox{yellow}{$74.05\pm 0.82$} & \colorbox{yellow}{$73.81\pm 1.26$} & $72.84\pm 1.47$ &$72.89\pm 0.93$& \colorbox{yellow}{$73.61\pm 1.36$} & \colorbox{yellow}{$73.90\pm 0.98$}
        \\
        $\delta$
        & \colorbox{yellow}{$73.85\pm 1.01$} & $72.37\pm 0.82$ & $72.05\pm 1.40$ &$71.93\pm 1.68$ & & 
        \\
        $\eta$ & \colorbox{yellow}{$73.37\pm 1.27$} & \colorbox{yellow}{$73.23\pm 1.73$} & \colorbox{yellow}{$74.36\pm 1.03$} & \colorbox{yellow}{$73.47\pm 0.86$} & & 
        \\
        $\zeta$ &  $72.99\pm 1.52$ & \colorbox{yellow}{$73.11\pm 1.80$} & $72.97\pm 1.26$ & $71.98\pm 0.66$ & & 
        \\
        $\omega$ & \colorbox{yellow}{$73.04\pm 1.57$} & \colorbox{yellow}{$74.36\pm 0.99$} & \colorbox{orange}{$74.62\pm 1.62$} & \colorbox{yellow}{$74.31\pm 1.87$} & &
        \\
        \hline\hline
    \end{tabular}
    \caption{Ablation study on the Citeseer dataset. The dictionary for the different terms in the CCNN architecture is indicated as in Table \ref{tab:options}.  Results highlighted in \colorbox{orange}{Orange} indicate the best performance while results indicated in \colorbox{yellow}{Yellow} are performances within one standard deviation of the best.}  
    \label{tab:ablation_Citeseer}
\end{table*}

\begin{table*}[]
    \centering
      \hspace*{-10mm}\begin{tabular}{||c|cccc|cc||}
    \hline\hline
         & $A+B+C$ & $B+C$ & $C$ &$A+C$& $B$ & $A+B$\\
        \hline
         $\gamma$
         &$85.26\pm 1.59$ & \colorbox{yellow}{$86.20\pm 0.82 $} & \colorbox{yellow}{$86.20\pm 0.89$} & $85.58\pm 0.87$ & \colorbox{yellow}{$86.56\pm 1.16$} & \colorbox{yellow}{$86.20\pm 0.18$} \\
        $\delta$
         & $85.70\pm 0.66$ &\colorbox{orange}{$86.62\pm 0.62$} & $84.11\pm 1.31$ & $83.96\pm 0.93$ & & 
         \\
         $\eta$ & \colorbox{yellow}{$86.38\pm 0.83$} & $85.41\pm 1.28$ & $85.32\pm 0.64$ & $84.93\pm 0.60$ & & 
         \\
         $\zeta$ & $85.47\pm 0.64$ & $85.61\pm 0.85$ & $84.34\pm 0.77$& $83.87\pm 0.84$ & & 
         \\
         $\omega$ & \colorbox{yellow}{$86.44\pm 0.61$} & \colorbox{yellow}{$86.17\pm 0.55$} & $85.82\pm 0.67$ & $85.23\pm 0.56$ & &\\
        \hline \hline
    \end{tabular}
    \caption{Ablation study  on the Cora dataset. The dictionary for the different terms in the CCNN architecture is indicated as in Table \ref{tab:options}.\colorbox{orange}{Orange} indicates the best performance and \colorbox{yellow}{Yellow} indicates performances within one standard deviation of the best.}
    \label{tab:ablation_Cora}
\end{table*}
\end{document}